\documentclass[12pt]{iopart}

\usepackage{graphicx}
\usepackage{cite}

\begin{document}
	\title[Reflectance of graphene-coated dielectric plates in the
framework of Dirac model]{Reflectance of graphene-coated dielectric
plates in the framework of Dirac model:
Joint action of energy gap and chemical potential}
	\author{G.~L.~Klimchitskaya${}^{1,2}$, V.~S.~Malyi${}^{2}$,
V.~M.~Mostepanenko${}^{1,2,3}$ and V.~M.~Petrov${}^{2}$}
	\address{${}^{1}$Central Astronomical Observatory at Pulkovo
		of the Russian Academy of Sciences, St.Petersburg, 196140, Russia}
	\address{${}^{2}$Institute of Physics, Nanotechnology and
		Telecommunications, Peter the Great Saint Petersburg
		Polytechnic University, St.Petersburg, 195251, Russia}
	\address{${}^{3}$Kazan Federal University, Kazan, 420008, Russia}
	
\ead{vmostepa@gmail.com}

	\begin{abstract}
We investigate the reflectance of a dielectric plate coated with
a graphene sheet which possesses the nonzero energy gap and chemical
potential at any temperature. The general formalism for the
reflectance using the polarization tensor is presented in the framework
of Dirac model. It allows calculation of the reflectivity properties
for any material plate coated with real graphene sheet on the basis of
first principles of quantum electrodynamics. Numerical computations of
the reflectance are performed for the graphene-coated SiO$_2$ plate at
room, liquid-nitrogen, and liquid-helium temperatures. We demonstrate
that there is a nontrivial interplay between the chemical potential,
energy gap, frequency, and temperature in their joint action on the
reflectance of a graphene-coated plate. Specifically, it is shown that
at the fixed frequency of an incident light an increase of the chemical
potential and the energy gap affect the reflectance in opposite
directions by increasing and decreasing it, respectively. According to
our results, the reflectance approaches unity at sufficiently low
frequencies and drops to that of an uncoated plate at high frequencies
for any values of the chemical potential and energy gap. The impact of
temperature on the reflectance is found to be more pronounced for graphene
coatings with smaller chemical potentials. The obtained results could
be applied for optimization of optical detectors and other devices exploiting
graphene coatings.
		\end{abstract}
\maketitle
	
	\section{\label{sec:level1}Introduction}
	
	Investigation of graphene, a two-dimensional sheet of carbon atoms packed in a hexagonal lattice,
has clearly demonstrated unique opportunities of this material for both fundamental and applied physics.
The Dirac model is of primary importance for understanding the physical properties of graphene.
According to this model, at energies below one or two~eV~the quasiparticles in graphene are either
massless or very light and obey the linear dispersion relation although move with the
Fermi velocity~$\upsilon_F$~which is less than the speed of light $c$ by approximately a factor
of 300 (see \cite{1,2,3}).
	
	Theoretical description of the electrical conductivity and reflectances of graphene has been
developed in the framework of the Kubo response theory, the Boltzmann transport equation, the
formalism of the correlation functions in the random phase approximation and some other,
more phenomenological approaches, such as the two-dimensional Drude model. This provided insight
into the nature of optical properties of graphene and offered a clearer view on its universal
conductivity expressed via the electric charge $e$ and the Planck constant $\hbar$
 \cite{4,5,6,7,8,9,10,11,12,13,14}. Some of the theoretical approaches mentioned above have been
 applied also to calculate the Casimir (Casimir-Polder) force between two graphene sheets and
 between an atom and graphene which is caused by the zero-point and themal fluctuations of the electromagnetic field ~\cite{15,16,17,18,19,20,21,22,22a,23,24,25,26,27}. It should be mentioned,
 however, that the cited theoretical approaches are not based on the first principles of quantum electrodynamics at nonzero temperature which should be directly applicable to a simple system
 described by the Dirac model. In this connection, due to some ambiguities in the statistical
 description and limiting transitions, several conflicting results related to the conductivity,
 optical properties and Casimir forces in graphene systems have been obtained
 (see a discussion \cite{28,29,30}).
	
	The most fundamental description of the optical properties, Casimir effect and electrical
conductivity of graphene is given by the polarization tensor in (2+1)-dimensional space-time.
Although in some specific cases this quantity was considered by many authors (see, e.g., \cite{31,32,33}), the exact expressions valid at any temperature for graphene
with nonzero energy gap have been derived only recently \cite{34,35}.
The application region of the obtained polarization tensor was, however, restricted to pure
imaginary Matsubara frequencies. Because of this, it was possible to apply it for detailed
investigation of the Casimir effect in graphene systems ~\cite{30,36,37,38,39,40,41,42,43,44,45},
but not for calculations of the reflectances and conductivities of graphene which are defined
at real frequencies.
	
	The polarization tensor valid for a gapped graphene over the entire plane of complex
frequencies, including the real frequency axis, was derived in \cite{46,46a}.
In \cite{47,47a} the results of \cite{46,46a} were generalized for the case of graphene
with nonzero chemical potential. The novel form of the polarization tensor
\cite{46,46a,47,47a} has been used for a more detailed study of the Casimir effect in
graphene systems ~\cite{48,49,50,51,52,53,54} and for investigation of the electrical
conductivity of graphene on the basis of first principles of quantum electrodynamics
at nonzero temperature ~\cite{55,56,57,58}. Special attention was devoted to the
field-theoretical description of the reflectivity properties of graphene.
In \cite{46,46a} and ~\cite{59} they were investigated in detail for both pure (gapless)
and gapped graphene, respectively, and in \cite{60} the joint action of nonzero
energy gap and chemical potential was analyzed.
	
	Of particular interest is the configuration of a dielectric plate coated with a
graphene sheet. In addition to the fundamental interest in this subject, graphene coatings
find prospective technological applications to optical detectors, various optoelectronic
devices, solar cells, biosensors  etc.~\cite{61,62,63,64}. Using the polarization tensor
of \cite{46,46a}, the reflectivity properties of material plates coated with pure~\cite{65,65a}
and gapped~\cite{66} graphene have been investigated. The case of thin films coated with
gapped graphene has been considered as well~\cite{67}. However, the joint action of nonzero
energy gap and chemical potential on the reflectivity properties of graphene-coated plates,
which is of major interest for condensed matter physics and its applications,
remains to date unexplored.
	
	In this paper, we present the formalism using the polarization tensor of \cite{46,46a,47,47a} and
calculate the reflectance of the dielectric plate coated with real graphene sheet which possesses
nonzero energy gap
in the spectrum of quasiparticles and chemical potential.
Thus, here we exploit the same formalism, as was used in Refs.~\cite{48,49,50,51,52,53,54}
in the Casimir configurations of two parallel graphene sheets (graphene-coated plates),
but in quite different physical situation.
All computations are performed at
the room temperature $T=300~$K, at the liquid-nitrogen temperature $T=77~$K, and at the
liquid-helium temperature $T=4.4~$K for the plate made of silica glass SiO$_2$. It is shown
that there is an interesting interplay between the chemical potential, energy gap, frequency
and temperature in their impact on the
reflectance of a graphene-coated plate.
	
	Another result obtained is that at the fixed frequency $\omega$ the increasing chemical
potential and energy
gap affect the reflectance in opposite directions. Specifically, an increase of the chemical potential
either increases the reflectance or leaves it unchanged. On the other hand, an increase of
the energy gap either decreases the reflectance or leaves it constant. It is shown that for
any values of the chemical potential and energy gap the reflectance of a graphene-coated plate
approaches unity at sufficiently
low frequencies. With increasing frequency, the reflectance drops to that of an uncoated plate.
According to our results, for larger chemical potential the frequency region, where the
reflectance of
a graphene-coated plate depends on the presence of coating, becomes wider. However,
at frequencies $\hbar\omega>20~$meV the graphene coating makes no impact on the reflectance
of a SiO$_2$ plate any
more. An impact of temperature on the reflectance is more pronounced for graphene
coating with
smaller chemical potentials. For sufficiently high chemical potential, temperature
no longer affects
the reflectance. By and large, the presented formalism allows computation of the
impact of graphene
coating on the reflectance of a plate made of any material.
	
	The paper is organized as follows. In Section 2 we present the formalism of the
Dirac model expressing
the reflectance of a graphene-coated plate via the polarization tensor.
Section 3 contains the computational results for the reflectance as a function of
frequency and chemical potential under the
varying energy gap and temperature. In Section 4 the reflectance is considered
as a function of energy
gap under different values of the chemical potential and temperature.
In Section 5 the reader will find our conclusions and a discussion.
	
	\section{\label{sec:level1}General formalism for the reflectance}
	
We consider sufficiently thick dielectric plate, which can be treated as a semispace,
coated with gapped graphene possessing some chemical potential $\mu$. It is well known
that for a pure freestanding graphene the energy gap in the spectrum of quasiparticles
$\Delta=0$, but under an influence of the defects of structure, electron-electron
interaction and in the presence of a material substrate the nonzero gap
$\Delta\leq 0.1$
or $0.2~$eV arises~\cite{1,3,33,67aa,67ab}.
As to the chemical potential, it describes the doping
concentration, which is nonzero for any real graphene sample and takes typical values
$\mu\sim 0.1~$eV~\cite{3}. Below we consider only the case of  normal incidence
so that the projection of the wave vector on a graphene plane $k=0$
(a dependence on the incident angle is discussed in \cite{60,65,66}).
Because of this, it is sufficient to consider only the transverse magnetic (TM), i.e.,
$p$-polarized, electromagnetic waves of frequency $\omega$ incident on a graphene-coated plate.
	
	The TM reflection coefficient on the dielectric plate coated with a graphene sheet is given by~\cite{45}
	\begin{equation}
	r_{\rm TM}^{(g,p)}(\omega,0)=
	\frac{r_{\rm TM}^{(g)}(\omega, 0)+r_{\rm TM}^{(p)}(\omega, 0)
[1-2r_{\rm TM}^{(g)}(\omega, 0)]}{1-r_{\rm TM}^{(g)}(\omega, 0)r_{\rm TM}^{(p)}(\omega, 0)}, \label{eq1}
	\end{equation}
	where $r_{\rm TM}^{(g)}$ and $r_{\rm TM}^{(p)}$ are the TM reflection coefficients on graphene and on a plate taken singly, respectively (see Appendix for more details).
The magnitude of the wave vector projection on the plane of graphene is put equal to zero because we consider the normal incidence.
	
	The TM reflection coefficient of the electromagnetic waves on a freestanding graphene sheet is expressed as ~\cite{46,67}
	\begin{equation}
	r_{\rm TM}^{(g)}(\omega,0)=\frac{\tilde{\Pi}_{00}(\omega,0)}
	{2+\tilde{\Pi}_{00}(\omega,0)}, \label{eq2}
	\end{equation}
	where the 00 component of the normalized polarization tensor
of graphene  $\tilde{\Pi}_{00}(\omega,0)$ is connected with the conventional definition
 of the 00 component of this tensor by~\cite{67}
	\begin{equation}
	\tilde{\Pi}_{00}(\omega,0)=-\frac{i\omega}{\hbar c} \lim_{k \to 0} \frac{\Pi_{00} (\omega,k)}{k^2}. \label{eq3}
	\end{equation}
	The explicit expression for the quantity $\Pi_{00}$ for graphene with nonzero $\Delta$ and $\mu$ at any temperature $T$ is presented in \cite{57} (see Appendix concerning different but equivalent expressions for the reflection coefficients on a freestanding graphene sheet in terms of the polarization tensor, electric susceptibility and density-density correlation functions).
	
	For us it is now important that the 00 component of the polarization tensor is directly connected with the in-plane conductivity of graphene ~\cite{44,55,56,57}
	\begin{equation}
	\Pi_{00} (\omega,k)=\frac{4 \pi i \hbar k^2}{\omega} \sigma (\omega,k). \label{eq4}
	\end{equation}	
	Combining this equation with (\ref{eq3}), one obtains
	\begin{equation}
	\tilde{\Pi}_{00}(\omega,0)=\frac{4 \pi }{c} \sigma (\omega,0). \label{eq5}
	\end{equation}
	As a result, the TM reflection coefficient on graphene (\ref{eq2}) is expressed in terms
of the in-plane conductivity
	\begin{equation}
	r_{\rm TM}^{(g)}(\omega,0)=\frac{2 \pi \sigma (\omega, 0)}{c+2 \pi \sigma (\omega, 0)}. \label{eq6}
	\end{equation}

	Now we return to the reflection coefficient $r_{\rm TM}^{(p)}$ on an uncoated dielectric plate. At the normal incidence it is given by the commonly known expression
	\begin{equation}
	r_{\rm TM}^{(p)}(\omega,0)=\frac{\sqrt{\varepsilon (\omega)}-1}{\sqrt{\varepsilon (\omega)}+1}=\frac{n(\omega)-1+ik(\omega)}{n(\omega)+1+ik(\omega)}, \label{eq7}
	\end{equation}
	where $\varepsilon (\omega)$ is the frequency-dependent dielectric permittivity of the plate material
and $n(\omega)$ and $k(\omega)$ are the real and imaginary parts of its complex index of
refraction, respectively. Representing the complex conductivity of graphene as
	\begin{equation}
	\sigma (\omega, 0)={\rm Re}\sigma (\omega)+i {\rm Im}\sigma (\omega, 0), \label{eq8}
	\end{equation}
	and substituting (\ref{eq6})--(\ref{eq8}) in (\ref{eq1}),
one finds the reflection coefficient on a graphene-coated plate
	\begin{equation}
	r_{\rm TM}^{(g,p)}(\omega,0)=
\frac{n(\omega)-1+\frac{4\pi}{c}{\rm Re}\sigma (\omega,0)+
i[k(\omega)+\frac{4\pi}{c} {\rm Im} \sigma (\omega,0)]} {n(\omega)+1+
\frac{4\pi}{c}{\rm Re}\sigma (\omega,0)+i[k(\omega)+
\frac{4\pi}{c} {\rm Im}\sigma (\omega,0)]}. \label{eq9}
	\end{equation}
{}	From (\ref{eq9}) one easily obtains the reflectance of the plate coated with a graphene sheet
	\begin{eqnarray}
&&
	{\cal R}(\omega)=[r_{\rm TM}^{(g,p)}(\omega,0)]^2
\label{eq10}	\\
&&
=\frac{[n(\omega)-1+\frac{4\pi}{c}{\rm Re}\sigma (\omega,0)]^2+
[k(\omega)+\frac{4\pi}{c} {\rm Im}\sigma (\omega,0)]^2} {[n(\omega)+1+
\frac{4\pi}{c}{\rm Re}\sigma (\omega,0)]^2+[k(\omega)+\frac{4\pi}{c} {\rm Im}\sigma (\omega,0)]^2}.
\nonumber	\end{eqnarray}
	
The exact expressions for real and imaginary parts of the in-plane conductivity of
graphene $\sigma(\omega,0)$ have been found in \cite{57} by using the polarization tensor.
For the real part it was obtained
\begin{equation}
\hspace*{-1cm}
{\rm Re}\sigma (\omega,0)=\sigma_0 \theta(\hbar \omega - \Delta)
	\frac{(\hbar \omega)^2+\Delta^2}{2(\hbar \omega)^2}
\left(\tanh\frac{\hbar \omega+2 \mu}{4 k_B T}+\tanh\frac{\hbar \omega-2 \mu}{4 k_B T}\right),
\label{eq11}
	\end{equation}
	where the universal conductivity of graphene is $\sigma_0={e^2}/(4 \hbar)$, the step function takes the values $\theta(x)=1$, $x\geq 0$ and $\theta(x)=0$, $x<0$, and $k_B$ is the Boltzmann constant.
	
	The imaginary part for the in-plane conductivity of graphene is somewhat more complicated~\cite{57}
	\begin{equation}
	{\rm Im}\sigma (\omega,0)=\frac{\sigma_0}{\pi}
\left[\frac{2 \Delta}{\hbar \omega}-\frac{(\hbar \omega)^2+\Delta^2}{(\hbar \omega)^2}
\ln\left|\frac{\hbar \omega +\Delta}{\hbar \omega -\Delta}\right|
+Y(\omega, \Delta, \mu)\right], \label{eq12}
	\end{equation}
	where the quantity $Y$ is defined as
	\begin{eqnarray}
&&
	Y(\omega, \Delta, \mu)=2\int_{\Delta/(\hbar \omega)}^\infty \!dt \sum_{\kappa=\pm 1}
\left(\exp\frac{\hbar \omega t+2\kappa \mu}{2k_BT}+1\right)^{-1}
\nonumber \\
&&~~~~~~~~~~~~~
\times\left[1+\frac{(\hbar \omega)^2+
\Delta^2}{(\hbar \omega)^2}\frac{1}{t^2-1}\right]. \label{eq13}
	\end{eqnarray}

Note that equations (\ref{eq11})--(\ref{eq13}) are exact in the framework of the Dirac model.
They are derived on the basis of first principles of quantum electrodynamics at nonzero
temperature. Taking into account that the Dirac model is applicable at sufficiently low
frequencies ($\hbar\omega<1-2~$eV), one arrives at the conclusion that these equations
give the proper description of all physical properties dealing with electrical conductivity
and reflection of light from graphene in this frequency region.
We remind that the imaginary part of the dielectric permittivity of graphene,  which is
proportional to the real part (\ref{eq11}) of its conductivity, is nonzero
over the wide frequency region.
This means that the concept of free Dirac quasiparticles allows the proper account of
energy losses by conduction electrons in graphene due to scattering processes
without resort to any additional parameters like in the 2-dimensional Drude model \cite{14}
or Kubo approach \cite{67j}.

Although physics behind
(\ref{eq11})--(\ref{eq13}) was already discussed in the literature \cite{57},
we briefly recall the meaning of different contributions to the conductivity.
Thus, the real part of conductivity (\ref{eq11}) is interpreted as originating from the
interband transitions \cite{57,67a}. The imaginary part of conductivity (\ref{eq12})
contains contributions from both the interband and intraband transitions.
To separate them analytically, one needs, however, to consider several distinct
asymptotic regions of parameters in the quantity $Y$ defind in (\ref{eq13}),
as is done in detail in \cite{57}. For one example, under the conditions
$k_BT\ll\mu$, $\hbar\omega\ll2\mu$, and $\Delta<2\mu$ one arrives at \cite{57}
\begin{equation}
{\rm Im}\sigma(\omega,0)=\frac{\sigma_0}{\pi}\left[
\vphantom{\frac{(\hbar \omega)^2+\Delta^2}{(\hbar \omega)^2}
\ln\left|\frac{\hbar \omega +\Delta}{\hbar \omega -\Delta}\right|}
\frac{4\mu}{\hbar\omega}-4\ln 2\frac{k_BT}{\hbar\omega}
-\frac{(\hbar \omega)^2+\Delta^2}{(\hbar \omega)^2}
\ln\frac{2\mu+\hbar \omega}{2\mu-\hbar \omega}\right].
\label{eq13a}
\end{equation}
\noindent
Here, the first two terms on the right-hand side correspond to the intraband transitions,
whereas the third term originates from the interband transitions \cite{7,57}
(see Ref.~\cite{57} for other relationships between $\hbar\omega$, $T$, $\mu$, and
$\Delta$ in (\ref{eq12}) and (\ref{eq13})).

	Using (\ref{eq10})--(\ref{eq13}) and the optical data for the complex index of
refraction of the plate material (see, for instance, \cite{68}), one can calculate
the reflectance of any material plate coated with real graphene sheet possessing nonzero
energy gap and chemical potential. Such computations are presented in the next sections.

	\section{\label{sec:level1}Reflectance as the function of frequency and chemical
potential at a varying energy gap}
	
	We apply the formalism of Sec.~II to describe the reflectance of a fused
	silica (SiO$_2$) plate coated with a graphene sheet. It has been known that
	fused silica plates are of frequent use as graphene
substrates \cite{69,70,71,72,73}.
	This material is characterized by the static dielectric permittivity
	$\varepsilon(0)\approx3.82$ which remains unchanged up to rather high frequency
	$\hbar \omega=10~$meV~\cite{68}. At higher frequencies the detailed optical
data for $n(\omega)$ and $k(\omega)$ for SiO$_2$ are contained in ~\cite{68}.
	
	Numerical computations of the reflectance of graphene-coated SiO$_2$
	plate are performed by using (\ref{eq10})--(\ref{eq13}). We start with the
dependence of reflectance on frequency for three different values of the chemical
potential and varying energy gap. The computational results are shown
	in figures~\ref{fig1}(a),~\ref{fig1}(b) and ~\ref{fig1}(c) at three temperatures $T=300~$K, $77$~K, $4.4$~K,
respectively.
	In each of the the gray regions from left
	to right are plotted for the values of the chemical potential
	$\mu=0$, $0.1$, and $0.8~$eV, respectively. The width of the gray regions
	is caused by the values of the energy gap which varies from
	$\Delta=0.1$~eV (the left boundary line of each region)
to $\Delta=0$ (the right boundary line of each region).
The bottom
	lines in ~\ref{fig1}(a),~\ref{fig1}(b) and ~\ref{fig1}(c), which are horizontal at
$\hbar\omega\leq 10~$meV, demonstrate the
	reflectance of an uncoated SiO$_2$ plate. It is obtained from
	~(10) by putting $\sigma(\omega,0)=0$
	\begin{equation}
	{\cal R}_0(\omega)=\frac{[n(\omega)-1]^2+k^2(\omega)}{[n(\omega)+1]^2+k^2(\omega)}. \label{eq14}
	\end{equation}
	In the frequency region  $\hbar\omega\leq 10~$meV
we have $k(\omega)=0$ \cite{68} and (\ref{eq14}) reduces to
	\begin{equation}
	{\cal R}_0(\omega)=
\left[\frac{\sqrt{\varepsilon(0)}-1}{\sqrt{\varepsilon(0)}+1}\right]^2
\approx0.1044. \label{eq15}
	\end{equation}
	
	As is seen in figure~\ref{fig1}, at the very low frequencies the reflectance of a
graphene-coated SiO$_2$ plate is close to unity and drops to the reflectance of an
uncoated plate with increasing frequency.
Physically this means that at low frequencies the graphene coating behaves like a metallic one,
whereas at high frequencies becomes transparent.
The frequency region, where this drop takes place,
shifts to higher frequencies with increasing chemical potential of graphene.
The width of the gray regions (i.e., the range of frequencies where the energy gap of
graphene sheet varies between zero and $0.1~$eV) quickly decreases with increasing $\mu$.
This means that for larger $\mu$ an impact of the energy gap on the reflectance quickly
decreases and for $\mu=0.8~$eV there is no impact of $\Delta$ on the reflectance at
any temperature. With decreasing temperature, some of the gray regions become wider.
This feature is especially pronounced for the undoped graphene ($\mu=0$).
Thus, at $T=77~$K and $4.4~$K the left boundaries of the left gray regions are missing
because they appear at much lower frequencies than those ones shown in figures~\ref{fig1}(b) and ~\ref{fig1}(c).
Note that for any reflectance smaller than unity the widths of all gray regions
increase with decreasing frequency, but decrease when the reflectance approaches
unity.
The right boundary lines of the left gray regions in figures~\ref{fig1}(a)
and \ref{fig1}(b), plotted for
the case $\mu=\Delta=0$ at $T=300~$K and $77~$K, coincide with the lines 3 and 2,
respectively, obtained earlier for this particular case in \cite{65}.

It is interesting that at a fixed frequency the increase of $\mu$ typically leads
to a larger (or the same) reflectance, whereas the increase of $\Delta$ results in a
smaller (or the same) reflectance, i.e., both quantities act in opposite directions.
This effect has simple physical explanation. Thus, increasing $\mu$ leads to an increased
density of charge carriers and increased conductivity resulting in the higher reflectance.
Just to the opposite, an increased energy gap leads to the decreased mobility of charge
carriers, decreased conductivity and, finally, to the lower reflectance.
At $\hbar \omega\geq 20~$meV the graphene coating does not make any impact on the
reflectance. Note that here and below we do not show the unobservable
infinitesimally narrow peaks ${\cal R}=1$ which appear under the condition
$\hbar \omega=\Delta$ (see \cite{66} for a discussion of this artefact).

	Now we consider a dependance of the reflectance on the chemical potential.
This is done separately at three different temperatures. In figure~\ref{fig2}, the
reflectance of a graphene-coated plate is shown as a function of $\mu$ at $T=300~$K.
The four gray regions from top to bottom are plotted for the frequency
$\hbar\omega=0.1$, $1$, $5$, and $10~$meV, respectively. In each of these regions,
the upper boundary line is plotted for the energy gap $\Delta=0$ and the lower
boundary line for $\Delta=0.1~$eV. The dashed lines are computed for $\Delta=0.05~$eV.
They are depictured only for sufficiently wide gray regions related to
$\hbar\omega=0.1$ and 1~meV. As is seen in figure~\ref{fig2}, with increasing
$\mu$ an impact of the energy gap on the reflectance quickly decreases in line with
figure~\ref{fig1}. The reflectance is the monotonously increasing function of $\mu$.
The maximum values of reflectance are reached faster for the light of lower frequency.
	
	Next, we present the computational results for the reflectance as a function of
$\mu$ at $T=77~$K. In figure~\ref{fig3} it is plotted (a) at $\hbar\omega=0.1$~meV and
(b) at $\hbar\omega=1.0$~meV. The upper and lower boundary lines of the gray regions
are plotted for the energy gap $\Delta=0$ and $0.1$~eV, respectively, whereas the
dashed middle lines for $\Delta=0.05~$eV. In an inset to figure~\ref{fig3}(b) the
region of small values of $\mu\leq 0.12~$eV is shown on an enlarged scale.
As can be seen in figure~\ref{fig3}, the reflectance is again a monotonously increasing
function of $\mu$. At $\omega=1.0~$meV [see figure~\ref{fig3}(b)] the lower boundary line
of the dashed region ($\Delta=0.1~$eV) is very similar to the respective line in
figure~\ref{fig2} plotted at $T=300~$K and differs from it by only a more sharp angle.
{}From figure~\ref{fig3}(a) it is seen that for $\mu\geq 0.12~$eV the reflectance
${\cal R}=1$ and does not depend on $\mu$. At $\hbar\omega=1.0~$meV the reflectance
approaches to unity at much larger $\mu\approx0.8~$eV [see figure~\ref{fig3}(b)].
We do not present
the computational results for a reflectance at $\hbar\omega=5~$and $10~$meV because
at $T=77~$K they are nearly the same as at $T=300~$K (see figure~\ref{fig2}).
	
	Finally, we repeat computations of the reflectance of a graphene-coated SiO$_2$
plate as a function of $\mu$ at $T=4.4~$K. The computational results are shown in
figures~\ref{fig4}(a) and \ref{fig4}(b) for the values of frequency $\hbar\omega=0.1~$ and $1.0$~meV,
respectively. All notations are the same as already explained in figures~\ref{fig3}(a) and \ref{fig3}(b).
It is seen that figure~\ref{fig4}(a) differs from figure~\ref{fig3}(a) plotted at $\omega=0.1~$meV,
$T=77~$K by a sharper angle of the lower line restricting the gray region.
It is seen also that at $\hbar\omega=0.1~$meV, $T=4.4~$K the graphene coating with
$\mu=0$~does not influence on the reflectance at any value of the energy gap from 0
to 0.1~eV (this is not the case at $\hbar\omega=0.1~$meV, $T=77~$K).
Figure~\ref{fig4}(b) plotted at $\hbar\omega=1.0~$meV, $T=4.4~$K differs from
figure~\ref{fig3}(b) plotted at the same frequency but $T=77~$K by only a sharper
angle of the lower line restricting the gray region (see the inset).
In this case at $\mu=0$ the graphene coating does not make an impact on the reflectance
at both $77~$K and $4.4~$K. At $T=4.4~$K the computational results for the reflectance
at the frequencies $\hbar\omega=5~$ and $10~$meV, are again nearly the same as
at $T=300~$K. They are shown by the bottom line and by the next to it gray region
in figure~\ref{fig2}.
	
	\section{\label{sec:level1}Reflectance as the function of energy gap under
 different chemical potentials and frequencies}

	In this section we present the computational results for the reflectance of
a graphene-coated SiO$_2$
plate as a function of the energy gap. In so doing, the chemical potential
$\mu$ may at times be fixed,
whereas the frequency takes several values, and at other times the fixed quantity
is the frequency and
the chemical potential takes different values. All computations are
again performed by using (\ref{eq10})--(\ref{eq13}).
In figure~\ref{fig5} the reflectance is shown as a function of
$\Delta$ for graphene coating with $\mu=0.01~$eV  at (a) $T=300~$K and
(b) $T=77~$K.
In figure~\ref{fig5}(a) the seven lines from top to bottom are plotted
for the values of
$\hbar\omega=~0.01$, $0.05$, $0.1$, $0.25$, $0.5$, $1$, and $\geq5$~meV,
 respectively.
In figure~\ref{fig5}(b) there are five lines for the values of
$\hbar\omega=0.01$, $0.05$, $0.1$, $0.5$
and $\geq 1$~eV. In both figures 5a and 5b the computational
results do not change when
$\hbar\omega$ becomes larger than $5$ and $1$~meV, respectively.
When decreasing temperature to below
$77$~K, the computational results do not change, so that figure~\ref{fig5}(b)
remains applicable.
As is seen in figure~\ref{fig5}, the reflectance decreases monotonously
with increasing $\Delta$.
At lower temperature this decrease goes on faster.
	
	In figure~\ref{fig6} similar computational results for the reflectance as
a function of $\Delta$ at $T=300~$K for graphene coating with $\mu=0.1~$eV.
It is interesting that for the relatively large $\mu$ the reflectance is
already temperature-independent, when decreasing $T$ to below $300$~K.
Because of this, the computational results at $T=77~$K and $4.4$~K are not shown.
The seven lines from top to bottom are plotted for $\hbar\omega=0.01$,
 $0.1$,  $0.5$, $1$, 2.5, 5, and 10~meV, respectively.
As is seen in figure~\ref{fig6}, in this case the reflectance varies with $\mu$ very slowly.
	
	Now we consider the dependence of the reflectance on energy gap at the fixed
frequency and different values of $\mu$. In figure~\ref{fig7} the reflectance of a
graphene-coated plate is shown as a function of the energy gap at
$\hbar\omega=0.1~$meV and (a) $T=300~$K, (b) $T=77~$K, and (c) $T=4.4~$K.
The seven lines from bottom to top are plotted for the values of chemical
potential $\mu=0$, $0.01$, $0.03$, $0.05$, $0.07$, $0.1$, and $\geq 0.5~$eV,
respectively. As is seen in figure~\ref{fig7}, the reflectance is again a decreasing
function of $\Delta$. At fixed $\Delta$ it takes either larger or the same values
for larger chemical potential. With decreasing temperature, a decrease of the
reflectance with increasing $\Delta$ becomes faster. At $T=4.4~$K [figure~\ref{fig7}(c)]
the graphene coating with $\mu=0$ does not influence on the reflectance shown by the bottom line.
At fixed $\Delta$ the impact of temperature on the reflectance is more pronounced for graphene
coating with smaller chemical potential. This is because at zero temperature the properties of
graphene with small $\mu$ are almost the same as for an undoped graphene.
	
	Finally, in figure~\ref{fig8} the computational results for the reflectance are shown
as the function of $\Delta$ at $\hbar\omega=1~$meV and (a) $T=300~$K, (b) $T=77~$K
(the values of the reflectance computed at $T=77~$K do not change with further decreasing
temperature). The seven lines from bottom to top are plotted for $\mu\leq 0.01~$eV,
$\mu=0.03$, $0.07$, $0.1$, $0.2$, $0.5$, and $0.8~$eV, respectively. As can be seen in
figure~\ref{fig8}, the top two lines corresponding to the largest values of the chemical
potential are nearly flat, i.e., the reflectance does not depend on $\Delta$ at any
temperature $T\leq 300$~K. It is seen also that the respective lines in figures~\ref{fig8}(a)
and ~\ref{fig8}(b) are very similar, i.e., the impact of temperature is very minor for
all considered values of the chemical potential.
	
	\section{\label{sec:level1}Conclusions and discussion}

	In this paper, we have presented the formalism, based of the first principles of
quantum electrodynamics at nonzero temperature, which allows to compute the reflectance of
a dielectric plate coated with real graphene sheet possessing nonzero energy gap and chemical
potential. In doing so, the graphene sheet is described by the polarization tensor in
(2+1)-dimensional space-time, and the dielectric plate by the measured optical data for
the complex index of refraction of the plate material. This formalism was applied to the
case of a graphene-coated plate made of silica glass at room, liquid-nitrogen, and
liquid-helium temperatures.
	
The results of numerical computations show that there is a nontrivial interplay in the
joint action of the energy gap, chemical potential, frequency of the incident light,
and temperature on the reflectance. According to our results, an increase of the chemical
potential and the energy gap of graphene coating influences the reflectance in opposite
directions by making it larger and smaller, respectively.
The physical explanation for this result is provided.
It is shown that at sufficiently
low frequencies the reflectance of a graphene-coated plate approaches unity, but goes
down to the reflectance of an uncoated plate at sufficiently
high frequencies.
In this respect graphene coating is qualitatively similar to a metallic film.
At $\hbar\omega\geq  20~$meV the graphene coating does not affect
the reflectance of a plate made of silica glass independently of the values of the energy
gap and chemical potential. Temperature makes major impact on the reflectance for graphene
coatings with relatively small chemical potential. For sufficiently high chemical potential,
a decrease from the room temp45erature down to the liquid-helium value does not influence
on the reflectance.
	
We underline that the graphene coating significantly
affects the reflectance only at frequencies
below approximately $\hbar\omega=20~$meV, i.e., fully in the application region of the
Dirac model which extends up to 1--2~eV. This means that the developed formalism can be
reliably used for controlling the reflectance of graphene-coated plates, e.g.,
for making it larger or smaller, by modifying the frequency, temperature, energy gap or
the doping concentration.

\begin{figure}[b]
\vspace*{-0cm}
\centerline{\hspace*{2.5cm}
\includegraphics{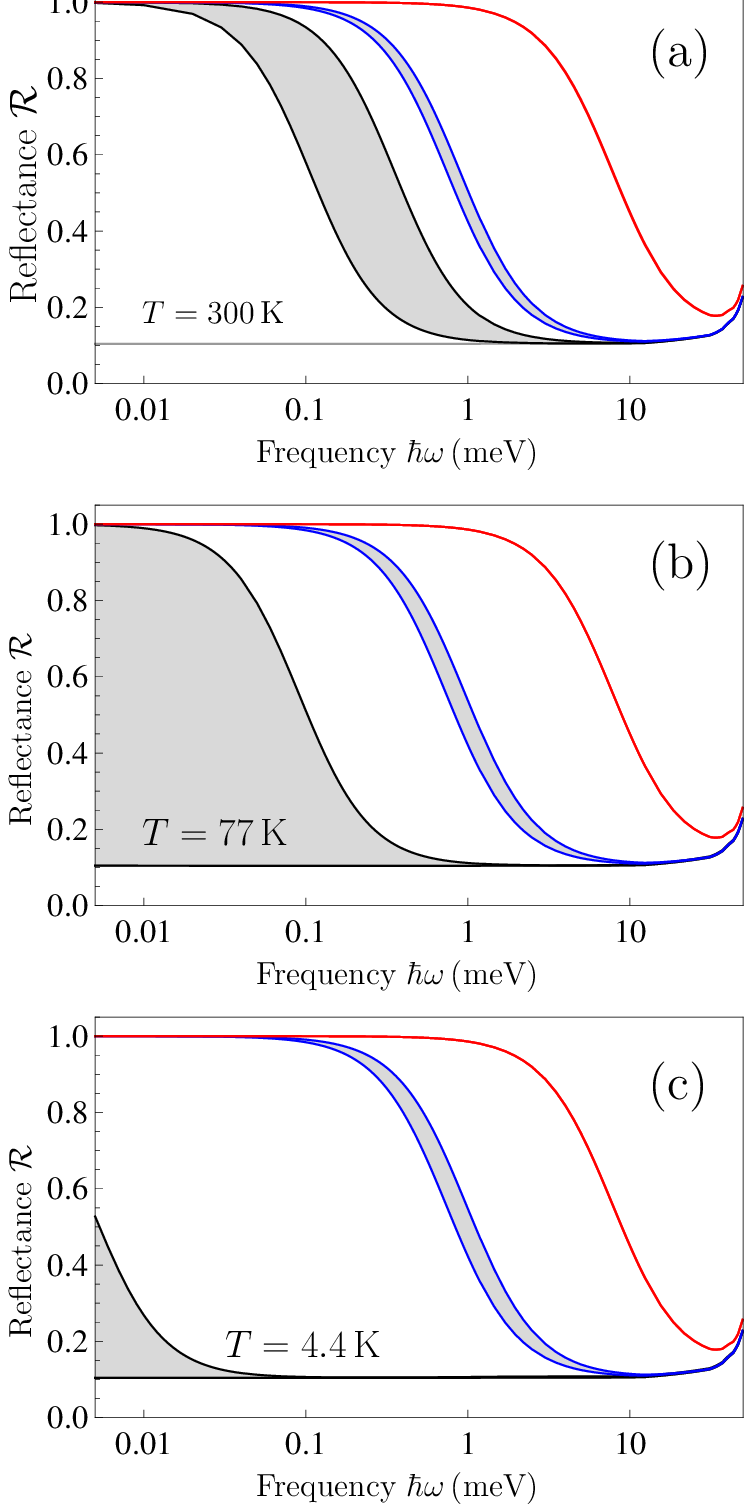}
}
\vspace*{-12.5cm}
\caption{\label{fig1}
The reflectances of graphene-coated SiO${}_2$ plate at
		 	the normal incidence are shown as functions of frequency at
		 	(a) $T=300~$K, (b) $T=77~$K, and (c) $T=4.4~$K. The gray
		 	regions from left to right are plotted for the chemical
		 	potential $\mu=0$, $0.1$, and $0.8~$eV, respectively. The
		 	right and left boundary lines of each gray region are plotted
		 	for the energy gap $\Delta=0$ and $0.1~$eV, respectively.
}
\end{figure}

\begin{figure}[b]
\vspace*{-5cm}
\centerline{\hspace*{2.5cm}
\includegraphics{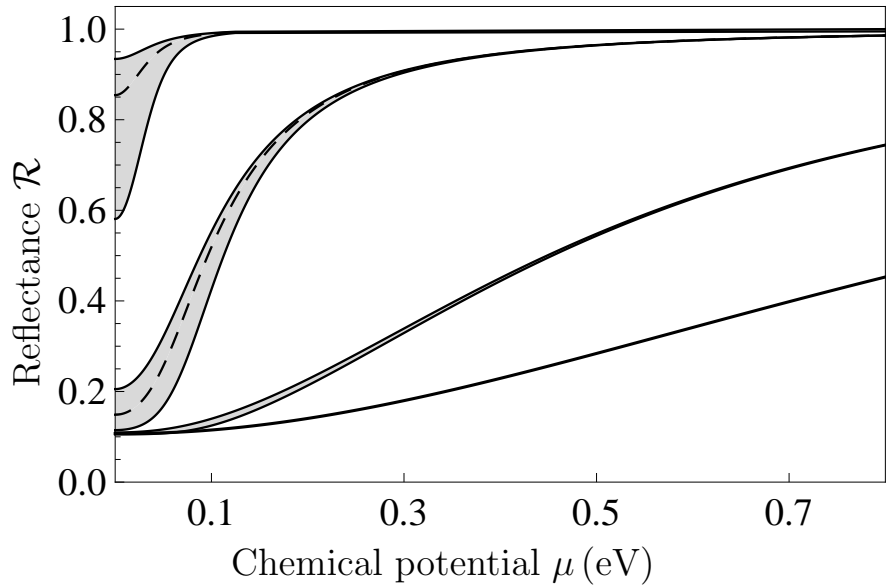}
}
\vspace*{-9.5cm}
\caption{\label{fig2}
The reflectances of graphene-coated SiO${}_2$ plate at
		 	the normal incidence are shown as functions of the chemical
		 	potential at $T=300~$K. The gray regions from top to bottom
		 	are plotted for the frequency $\hbar\omega=0.1$, $1$, $5$, and
		 	$10~$meV, respectively. The upper and lower boundary lines of
		 	each gray region are plotted for the energy gap $\Delta=0$
	 	and $0.1~$eV, respectively, whereas the dashed
middle lines are depictured for
		 	$\Delta=0.05~$eV.
}
\end{figure}

\begin{figure}[b]
\vspace*{1cm}
\centerline{\hspace*{2.5cm}
\includegraphics{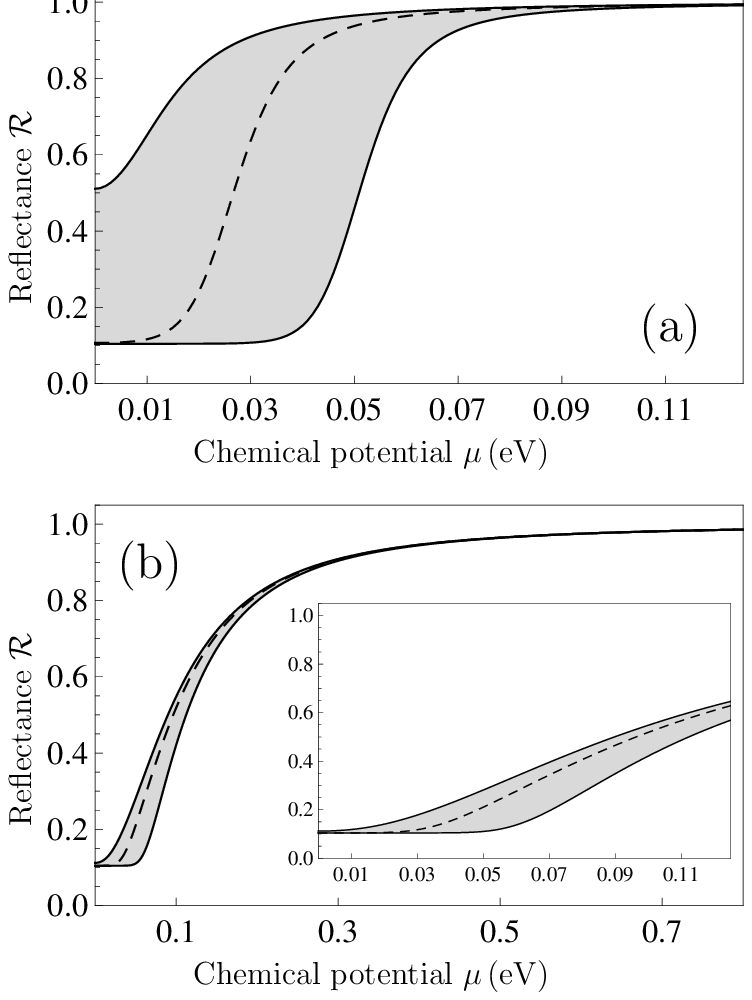}
}
\vspace*{-16.5cm}
\caption{\label{fig3}
The reflectances of graphene-coated SiO${}_2$ plate at
		 	the normal incidence are shown as functions of the chemical
		 	potential at $T=77~$K and (a) $\hbar\omega=0.1~$meV,
(b) $\hbar\omega=1.0~$meV.
		 	The upper and lower boundary lines of
		 	the gray regions are plotted for the energy gap $\Delta=0$
		 	and $0.1~$eV, respectively, whereas the dashed middle lines
		 	for $\Delta=0.05~$eV. The region of smaller $\mu$ is shown on
		 	an inset for $\hbar\omega=1.0~$meV.
}
\end{figure}

\begin{figure}[b]
\vspace*{+1cm}
\centerline{\hspace*{2.5cm}
\includegraphics{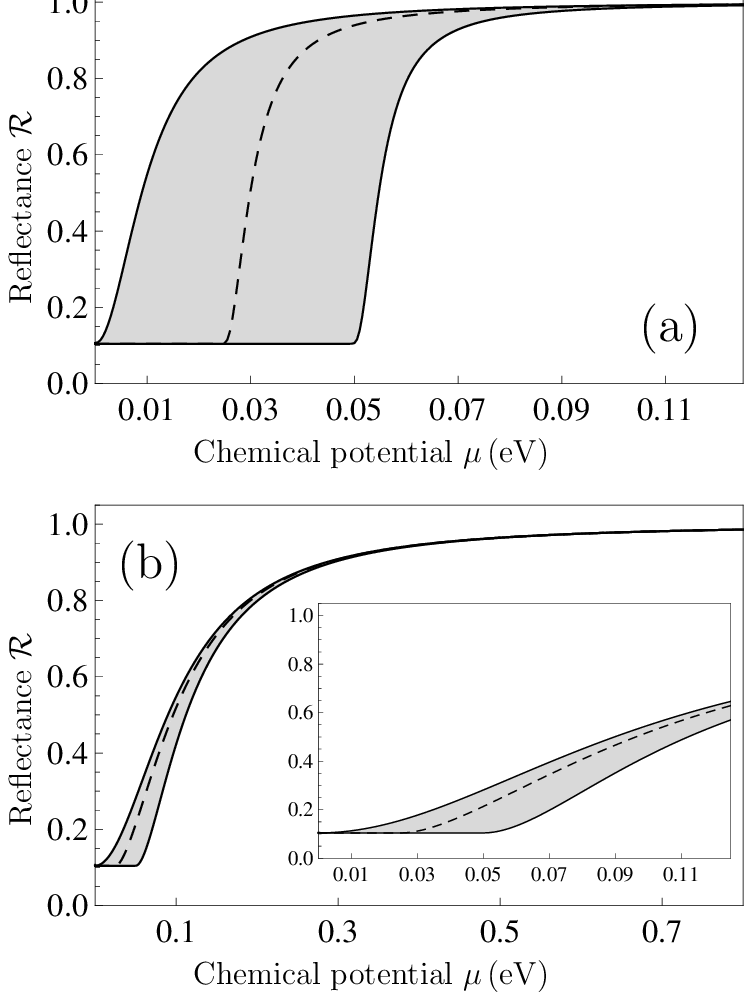}
}
\vspace*{-15.5cm}
\caption{\label{fig4}
The reflectances of graphene-coated SiO${}_2$ plate at
		 	the normal incidence are shown as functions of the chemical
		 	potential at $T=4.4~$K and (a) $\hbar\omega=0.1~$meV,
		 	(b) $\hbar\omega=1.0~$meV. The upper and lower boundary lines of
		 	the gray regions are plotted for the energy gap $\Delta=0$
		 	and $0.1~$eV, respectively, whereas the dashed middle lines
		 	for $\Delta=0.05~$eV. The region of smaller $\mu$ is shown on
		 	an inset for $\hbar\omega=1.0~$meV.
}
\end{figure}

\begin{figure}[b]
\vspace*{+1cm}
\centerline{\hspace*{2.5cm}
\includegraphics{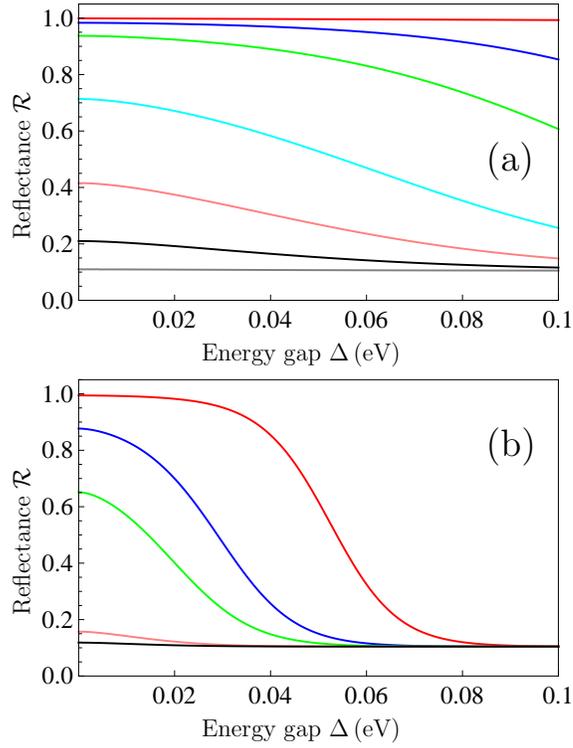}
}
\vspace*{-15.5cm}
\caption{\label{fig5}
The reflectances of graphene-coated SiO${}_2$ plate at
		 	the normal incidence are shown as functions of the energy gap
		 	for the chemical potential $\mu=0.01~$eV and at (a) $T=300~$K,
		 	(b) $T=77~$K. The lines from top to bottom are plotted for (a)
		 	$\hbar\omega=0.01$, $0.05$, $0.1$, $0.25$, $0.5$, $1.0$, and
		 	$\geq5$~meV, and (b) $\hbar\omega=0.01$, $0.05$, $0.1$, $0.5$,
		 	and $\geq1.0$~meV, respectively.
}
\end{figure}

\begin{figure}[b]
\vspace*{-4cm}
\centerline{\hspace*{2.5cm}
\includegraphics{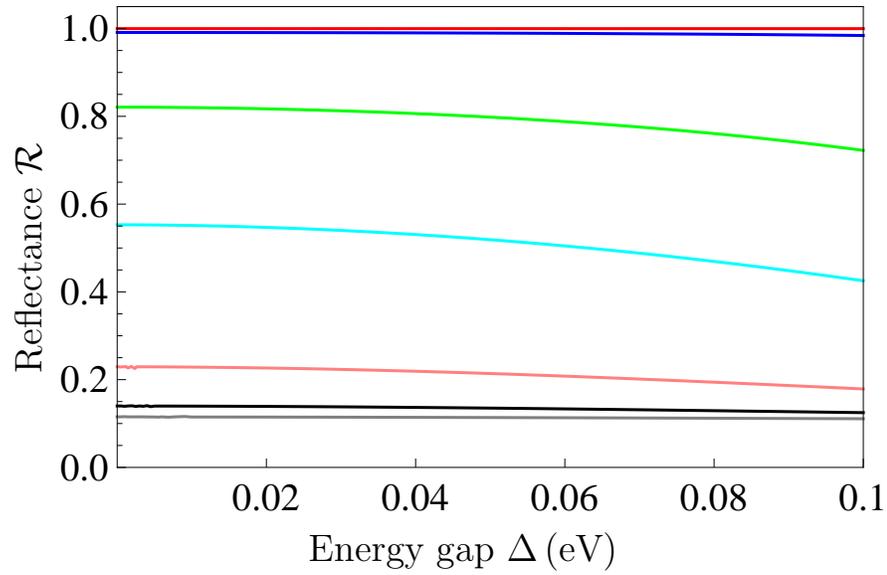}
}
\vspace*{-9.5cm}
\caption{\label{fig6}
The reflectance of graphene-coated SiO${}_2$ plate at
		 	the normal incidence is shown as a function of the energy gap
		 	for the chemical potential $\mu=0.1~$eV at $T=300~$K.
		 	The seven lines from top to bottom are plotted for
		 	$\hbar\omega=0.01$, $0.1$, $0.5$, 1.0, $2.5$, $5$, and
		 	$10$~meV.
}
\end{figure}

\begin{figure}[b]
\vspace*{-0cm}
\centerline{\hspace*{2.5cm}
\includegraphics{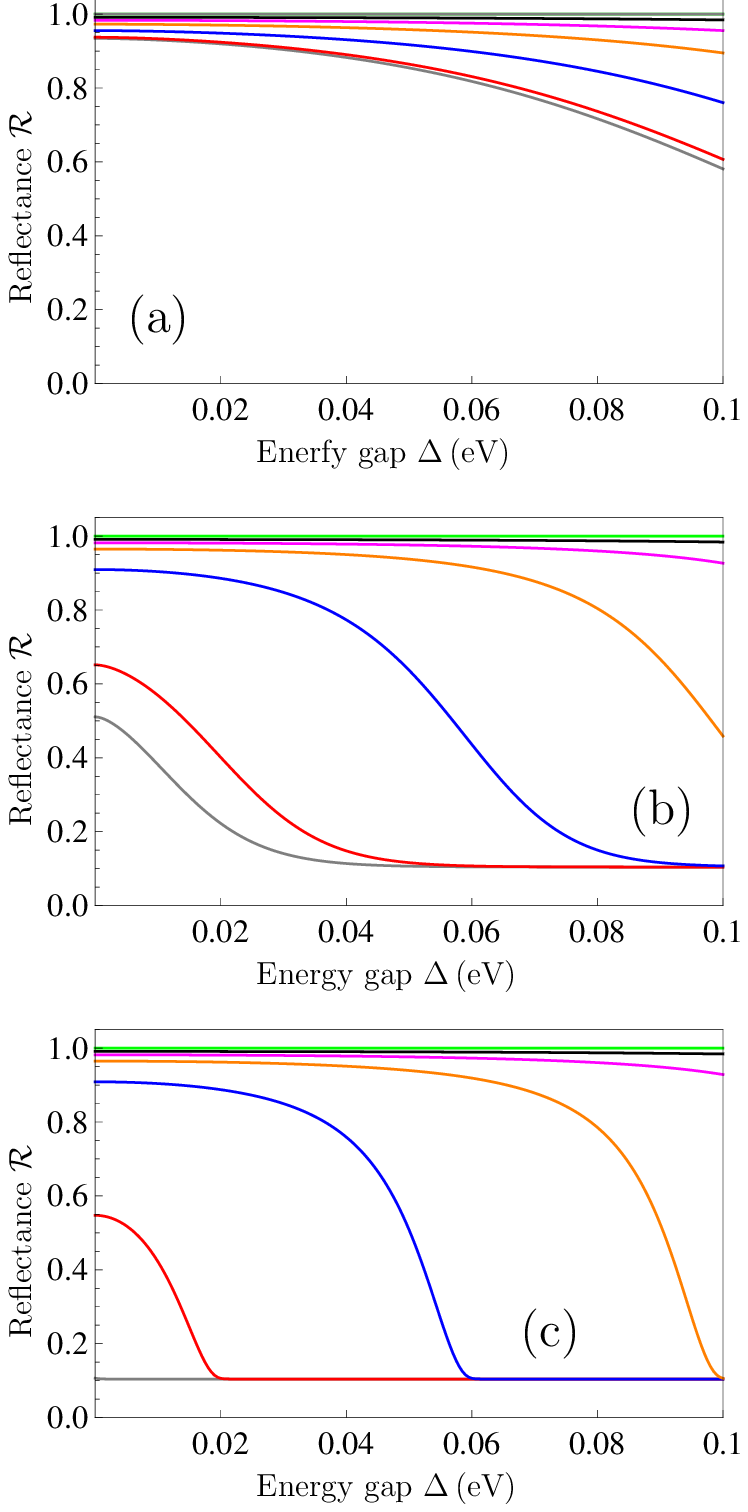}
}
\vspace*{-12.5cm}
\caption{\label{fig7}
The reflectances of graphene-coated SiO${}_2$ plate at
		 	the normal incidence are shown as functions of the energy gap
		 	for $\hbar\omega=0.1~$eV at (a) $T=300~$K, (b) $T=77~$K,
		 	and (c) $T=4.4~$K. The seven lines from bottom to top are
		 	plotted for the chemical potential $\mu=0$, $0.01$, $0.03$,
		 	$0.05$, $0.07$, $0.1$, and $\geq0.5$~eV, respectively.
}
\end{figure}

\begin{figure}[b]
\vspace*{-0cm}
\centerline{\hspace*{2.5cm}
\includegraphics{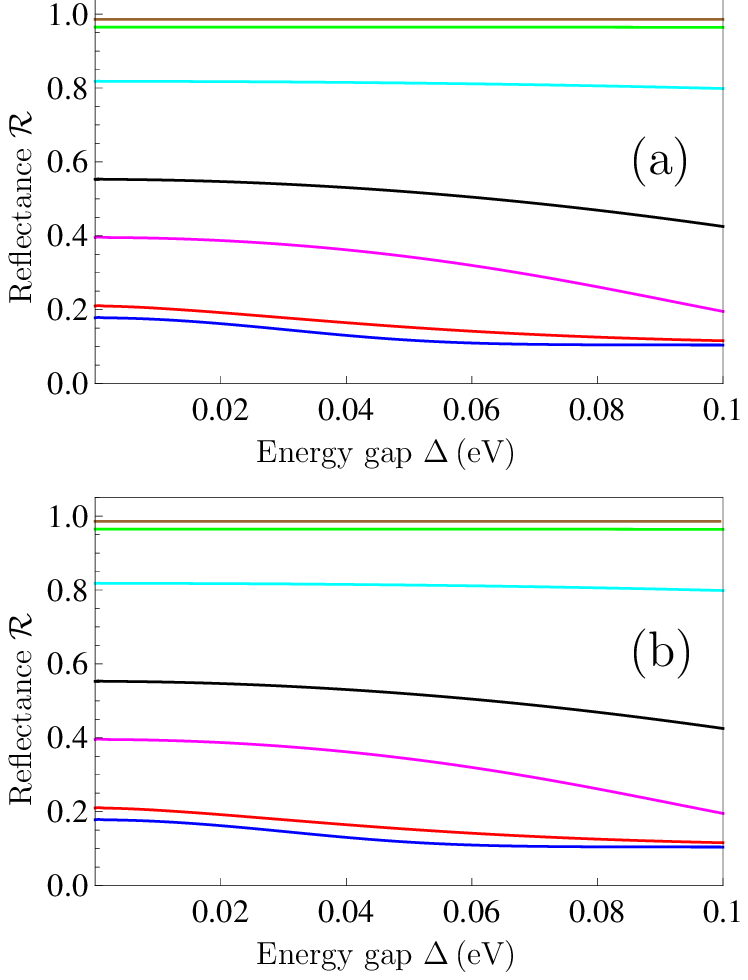}
}
\vspace*{-17.cm}
\caption{\label{fig8}
The reflectances of graphene-coated SiO${}_2$ plate at
		 	the normal incidence are shown as functions of the energy gap
		 	for $\hbar\omega=1~$meV at (a) $T=300~$K and (b) $T=77~$K.
		 	The seven lines from bottom to top are plotted for the chemical
		 	potential $\mu\leq0.01~$eV, $\mu=0.03$, $0.07$, $0.1$,
		 	$0.2$, $0.5$, and $0.8~$eV, respectively.
}
\end{figure}

It would be of much interest to compare the predicted effects with experimental data.
Currently, however, available experiments are performed at higher frequencies.
Thus, the reflectance of graphene-coated SiO${}_2$ substrates was measured at
$\hbar\omega$ above 60~meV \cite{84}, 130~meV \cite{85}, and 500~meV \cite{67a}.
In these frequency regions, an impact of graphene on the reflectance of a substrate
is rather small and
was found in agreement with calculations using the polarization tensor with
$\Delta=\mu=0$ \cite{65,65a} (note  that with increasing frequency the influence of
nonzero $\Delta$ and $\mu$ on the graphene reflectance decreases).
The relative infrared reflectance of graphene deposited
on a SiC substrate, investigated for different values of $\mu$ at
$\hbar\omega\geq 12~$meV \cite{86}, is in qualitative agreement with  the computational
results presented in Fig.~\ref{fig1}.
We conclude that large impact of graphene coating on a substrate reflectance in
the far infrared region and
the possibilities to control it predicted above might be beneficial
for numerous applications of graphene-coated plates mentioned in section 1.
	
	\ack{V.~M.~M. was partially funded by the Russian Foundation for Basic
Research Grant No. 19-02-00453 A.      		
The work of V.~M.~M. was also partially supported by the Russian Government Program of
Competitive Growth of Kazan Federal University.}

{\appendix
\section*{Appendix}
\setcounter{section}{1}

In this Appendix, we derive the basic expression (\ref{eq1}) for the reflection coefficient of
a graphene-coated plate \cite{45}. We also consider how the reflection coefficients of a
graphene sheet are expressed via the polarization tensor of graphene and other related
quantities such as the electric susceptibility and the density-density correlation function.
This helps to understand equations (\ref{eq2})--(\ref{eq4}).

Let us consider the system consisting of a graphene sheet, characterized by the amplitude
reflection coefficient $r^{(g)}(\omega,k)$ and the transmission coefficient $t^{(g)}(\omega,k)$
spaced in vacuum at a hight $a$ above thick dielectric plate characterized by the amplitude
reflection coefficien $r^{(p)}(\omega,k)$. These coefficients can be either transverse
magnetic (TM) or transverse electric (TE).
With account of multiple reflections on the graphene sheet and on the top boundary plane
of the dielectric plate, the reflection coefficient of our system takes the form \cite{82}
\begin{eqnarray}
&&
r^{(g,p)}(\omega,k)=r^{(g)}(\omega,k)+t^{(g)}(\omega,k)r^{(p)}(\omega,k)t^{(g)}(\omega,k)
e^{2iap(\omega,k)}
\nonumber\\
&&~~~~~~~~
\times
\sum_{n=0}^{\infty}\left[r^{(g)}(\omega,k)r^{(p)}(\omega,k)e^{2iap(\omega,k)}\right]^n
\nonumber\\
&&~~~
=r^{(g)}(\omega,k)+
\frac{t^{(g)}(\omega,k)r^{(p)}(\omega,k)t^{(g)}(\omega,k)
e^{2iap(\omega,k)}}{1-r^{(g)}(\omega,k)r^{(p)}(\omega,k)e^{2iap(\omega,k)}},
\label{eqA1}
\end{eqnarray}
\noindent
where $p(\omega,k)=(\omega^2/c^2-k^2)^{1/2}$.

Taking into account that for a graphene sheet in vacuum in the case of TM polarization
of the electromagnetic field one has $t^{(g)}(\omega,k)=1-r^{(g)}(\omega,k)$ \cite{83},
(\ref{eqA1}) can be rewritten in the form
\begin{equation}
r_{a,\rm TM}^{(g,p)}(\omega,k)=
\frac{r_{\rm TM}^{(g)}(\omega,k)+r_{\rm TM}^{(p)}(\omega,k)[1-2r_{\rm TM}^{(g)}(\omega,k)]
e^{2iap(\omega,k)}}{1-r_{\rm TM}^{(g)}(\omega,k)r_{\rm TM}^{(p)}(\omega,k)e^{2iap(\omega,k)}},
\label{eqA2}
\end{equation}

By considering the limiting case $a\to 0$ in (\ref{eqA2}), one obtains the reflection
coefficient \cite{45}
\begin{equation}
r_{0,\rm TM}^{(g,p)}(\omega,k)\equiv r_{\rm TM}^{(g,p)}(\omega,k)
=
\frac{r_{\rm TM}^{(g)}(\omega,k)+r_{\rm TM}^{(p)}(\omega,k)[1-
2r_{\rm TM}^{(g)}(\omega,k)]}{1-r_{\rm TM}^{(g)}(\omega,k)r_{\rm TM}^{(p)}(\omega,k)}
\label{eqA2a},
\end{equation}
\noindent
which coincides with (\ref{eq1}) at the normal incidence $k=0$.

The TM reflection coefficient of the electromagnetic waves on a graphene sheet
can be written in the form \cite{4,6,21}
\begin{equation}
r_{\rm TM}^{(g)}(\omega,k)
=
\frac{p(\omega,k)\alpha(\omega,k)}{ik+p(\omega,k)\alpha(\omega,k)},
\label{eqA3}
\end{equation}
\noindent
where the longitudinal (in-plane) electric susceptibility (polarizability) of graphene
$\alpha(\omega,k)$ is expressed via the respective dielectric permittivity and
density-density correlation function as
\begin{equation}
\alpha(\omega,k)=\varepsilon(\omega,k)-1=
-\frac{2\pi e^2}{k}\chi(\omega,k).
\label{eqA4}
\end{equation}
\noindent
The latter is directly connected with the in-plane conductivity of graphene \cite{21}
\begin{equation}
\sigma(\omega,k)=\frac{i e^2\omega}{k^2}\chi(\omega,k).
\label{eqA5}
\end{equation}
\noindent
Note that all the quantities $\alpha$, $\varepsilon$, $\chi$, and $\sigma$ are also
temperature-dependent.

The longitudinal (in-plane) density-density correlation function of graphene is expressed
via the component $\Pi_{00}$ of the polarization tensor \cite{44}
\begin{equation}
\chi(\omega,k)=-\frac{1}{4\pi  e^2\hbar}\Pi_{00}(\omega,k).
\label{eqA6}
\end{equation}
\noindent
In the framework of the Dirac model, the polarization tensor represents the effective
action for massless (or very light) fermionic quasiparticles in the quadratic order of
external electromagnetic field.

Substituting (\ref{eqA4}) and  (\ref{eqA6}) in  (\ref{eqA3}), one obtains the TM
reflection coefficient on a graphene sheet in terms of the polarization tensor
\begin{equation}
r_{0,\rm TM}^{(g)}(\omega,k)
=
\frac{p(\omega,k)\Pi_{00}(\omega,k)}{2i\hbar k^2+p(\omega,k)\Pi_{00}(\omega,k)},
\label{eqA7}
\end{equation}
\noindent
Introducing here the notation (\ref{eq3}), one arrives at (\ref{eq2}).

Then, by combining (\ref{eqA5}) with (\ref{eqA6}), we obtain equation (\ref{eq4}).
}



\end{document}